\documentstyle[prl,aps,amssymb,graphicx]{revtex}

\newcommand{\meV}{\mathrm{meV}}
\newcommand{\mV}{\mathrm{mV}}
\newcommand{\nm}{\mathrm{nm}}
\begin{document}
\twocolumn[
\hsize\textwidth\columnwidth\hsize\csname@twocolumnfalse\endcsname
\title{Shell filling in non-linear magneto-tunneling spectroscopy of vertical quantum dots}
\author{B.~Jouault$^{1}$, G.~Faini~\footnote{}$^{
1}$, A.~Angelucci$^{2}$, M.~Di~Stasio$^{2}$, G.~Santoro$^{3}$, A.~Tagliacozzo$^{2}$, F.~Laruelle$^{1}$, R.~Werner$^{4}$, A.~Forchel$^{4}$}
\address{
$^{1}$ L2M-CNRS, 196 Avenue H. Rav\'era, BP107, F-92225 Bagneux Cedex, France \\
$^{2}$ INFM and Universit\`a di Napoli ``Federico II'', 
Mostra d'Oltremare, Pad. 19, I-80125 Napoli, Italy \\
$^{3}$ S.I.S.S.A. and INFM, Via Beirut 2-4, I-34013 Trieste, Italy \\
$^{4}$ Technische Physik, Universit\"at W\"urzburg, Am Hubland, D-97074 W\"urzburg, Germany  \\
$*$e-mail:giancarlo.faini@L2M.CNRS.fr}

\maketitle
 \begin{abstract}
We report on non-linear magneto-tunneling experiments carried out in single GaAs vertical quantum dots. We show that conduction at low voltage bias  can be 
a spectroscopic tool for both the ground state and first excited states of few electrons on the dot. Increasing voltage a large resonant peak is observed and attributed to tunneling across the quasi-continuum of higher excited states.
\end{abstract}

]

Vertical quantum dots
 (QDs) obtained on double barrier resonant tunneling heterostructures better 
 than those realized on lateral 2DEG systems 
allow to investigate  the few electrons limit by starting from an empty dot and
progressively charging it. However linear transport experiments require the addition of a lateral gate around the dots: the fabrication of such devices is a real technical difficulty and only  few realizations have been reported 
in the literature till now~\cite{kou97}.

We report here on new results in  electron spectroscopy  of more conventional two-contacted vertical QDs without a lateral gate. These structures are relatively easier  to fabricate , but the theoretical analysis is difficult because  transport is non-linear.

Coulomb staircase  has been observed in these structures up to now only in presence of a strong barrier-thickness asymmetry~\cite{sch95,sle96}:
 when the electrons enter the QD from the thin barrier side,
 charge accumulation with subsequent electron-electron interaction is enhanced.
 {\it Viceversa, } one-electron states have been observed when electrons enter the QD from  the thick barrier side~\cite{bo92}.
We have used devices with symmetric barriers and, as discussed in the following, the conductance pattern at low bias can be attributed both to the  ground states and to  the excited states for N-electrons $( N\leq 6 )$.
At larger voltage bias a crossover to a quasi continuum of excited states in the dot produces a huge resonant peak as in diodes which 
are not laterally confined. 
 
The double barrier structures were grown by molecular beam epitaxy on n+ GaAs substrate. The undoped active layers consist of a 5.1nm quantum well sandwiched between 5nm $Al_{0.33}Ga_{0.67}As$ barriers. Two 500nm GaAs layers, which are n-doped with Si to $10^{18}\mathrm{cm}^{-3}$, provide the top and bottom contacts to the double barrier structure. The barriers are separated from the contacts by  undoped spacer layers to prevent Si segregation in the QD. Circular pillars of radii varying between 50nm and 1$\mu \mathrm{m}$ were fabricated using electron-beam lithography and $Ar/Cl_{2}$ reactive ion etching. A polyimide planarisation patterning was used for the bonding pads~\cite{fai91}.  
 
\begin{figure}
\begin{minipage}[c]{0.25 \textwidth}
\includegraphics*[width=1.5in]{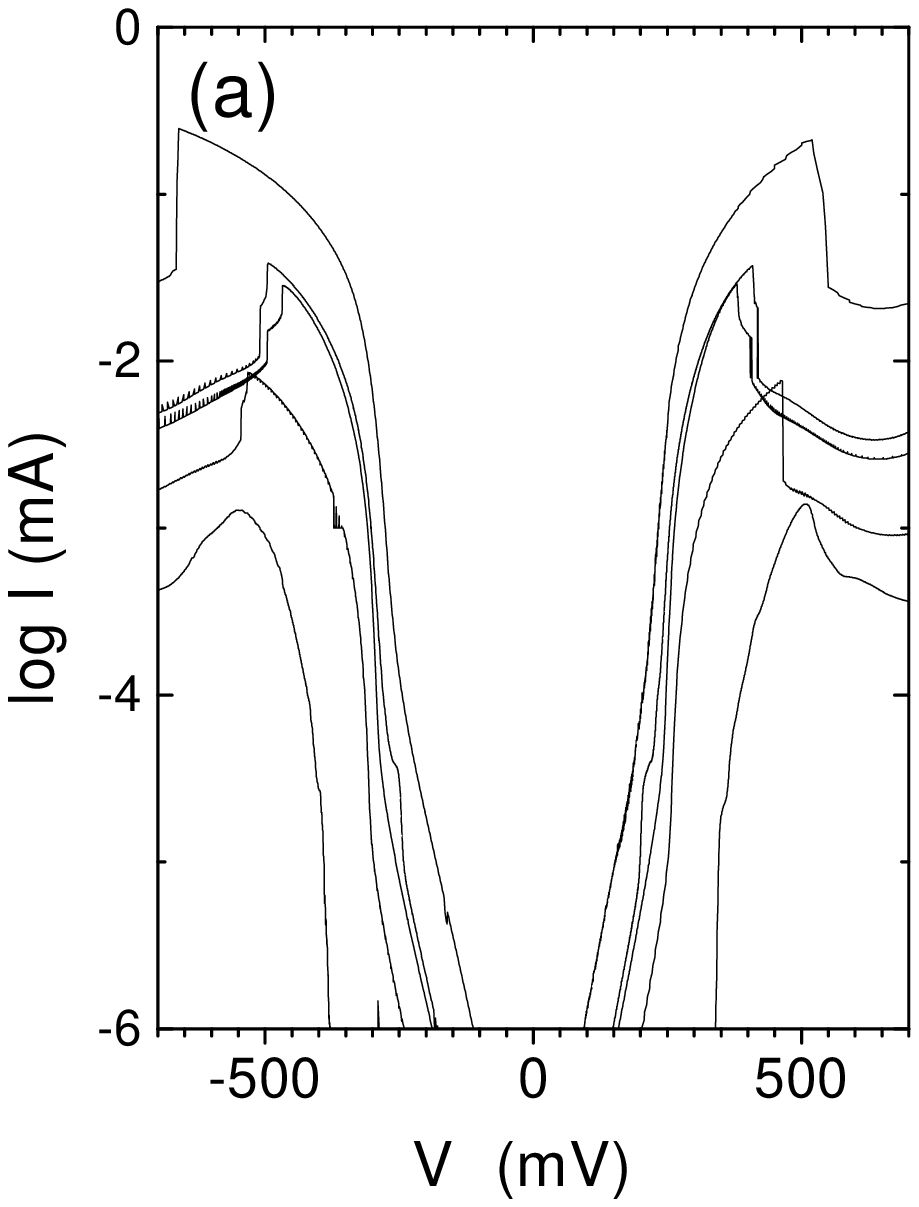}
\end{minipage}%
\begin{minipage}[c]{0.25 \textwidth}
\includegraphics*[width=1.5in]{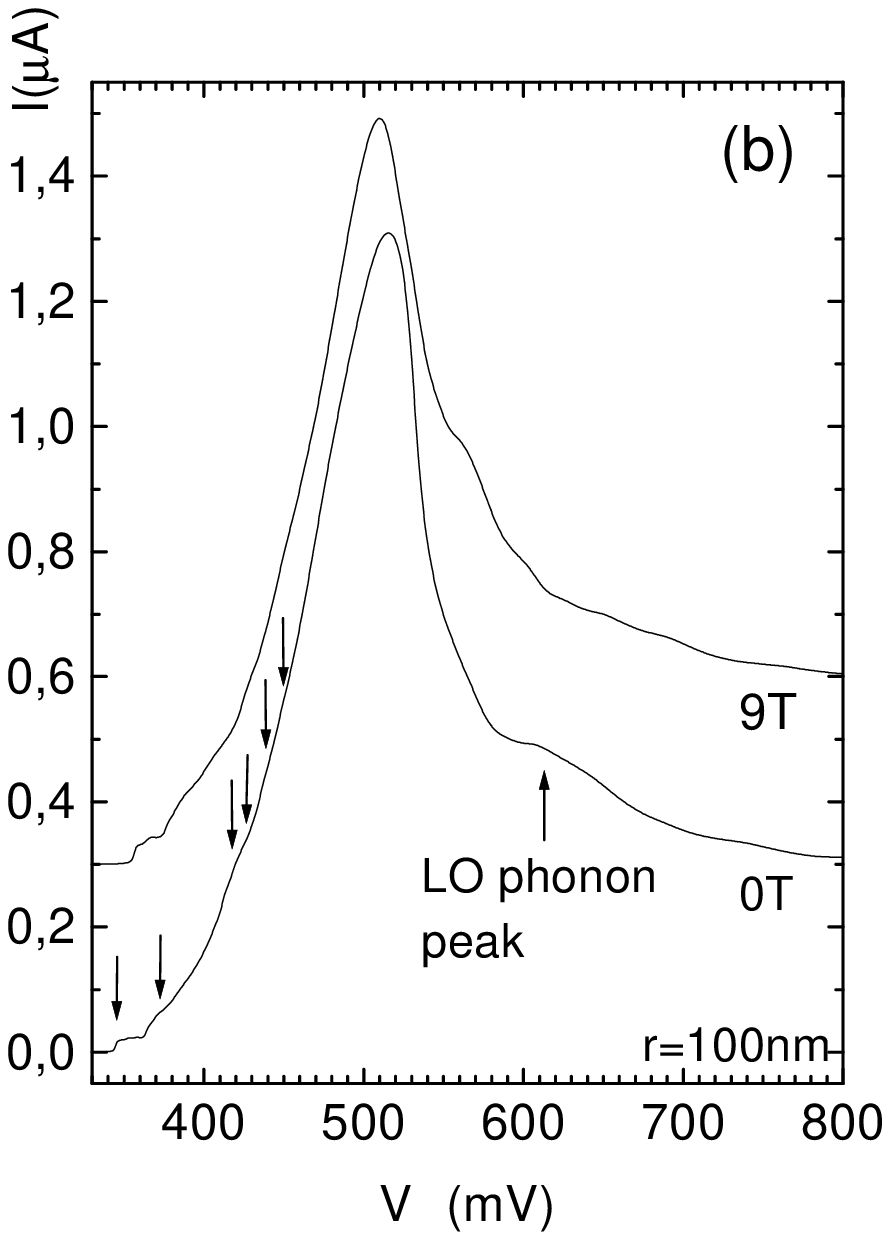}
\end{minipage}
\caption{\protect{\bf{(a)}} I-V curves at $T=4K$ for different pillars: curves refer to increasing radii (100,200,300, 350 and 1000 nm), from bottom to top; 
\protect{\bf{(b)}} I-V curve of the smallest pillar ($r=100 \nm$).}
\label{fig:iv}
\end{figure}

Fig.~\ref{fig:iv}a shows the I-V characteristics of diodes as a function of the lateral size, at a temperature of $4K$. The voltage threshold shifts to higher bias when the radius decreases 
and the current density decreases as well.  This reflects the growing influence of the lateral depletion length.
In the following we will focus only on the experimental data collected on the smallest pillar whose radius is 100nm. A magnetic field is applied parallel to the current flow. Fig.~\ref{fig:iv}b shows the I-V curves at $B=0T$ and $9T$ at a temperature of 4K. 
A clear staircase-like structure appears in the low voltage bias region, whose last steps are poorly resolved because of the sharp increase of the current.
Similar staircase structures have been widely reported by different groups and originate either from impurities or from QD's  created by lateral confinement.
 Whenever a localized state is aligned with the emitter Fermi level, resonant tunneling is allowed and a plateau appears in the I-V characteristics.
We argue in the following that, in our case,   the plateaus are due to 
the discreteness of the $N-$electron states in  the QD.

\begin{figure}
\includegraphics*[width=\linewidth]{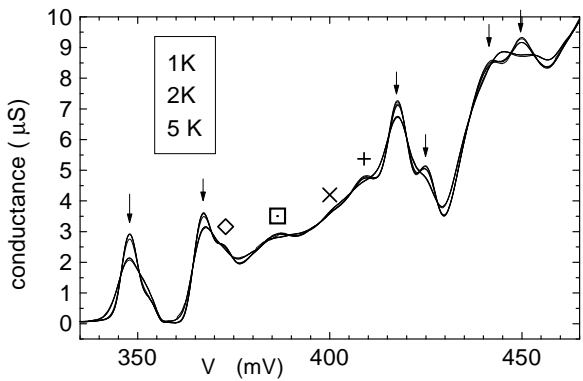}
\caption{Spectra of  $dI/dV$ vs $V$  at $T=1,2$ and $5K$. } 
\label{fig:tdep}
\end{figure}

Fig.~\ref{fig:tdep} shows the dynamical conductance $dI/dV$ of the pillar as a function of the voltage bias at three temperatures ranging from $T=1K$ to $T=5K$. 
From the temperature dependence of all the peaks labeled by $\downarrow$,$\lozenge$,$\boxdot$,$\times$, and $+$  in the figure  we conclude that  they correspond to the tunneling of electrons through QD states at the emitter Fermi level.
 In fact, taking into account the intrinsic width of the QD states, we fit the width at half maximum of the first conductance peak by using the formula:
\begin{equation}
dI/dV \propto \int_{-\infty}^{+\infty}
F(E-E_{F})
\frac{d}{dV}\left[T(E- \alpha V)\right]   dE
\end{equation}
where $F(E)$ is the Fermi distribution, $E_{F}$ is the Fermi energy in the emitter,  $\alpha$ is the energy-to-voltage conversion factor, and $T(E)$ is a transmission probability with a Lorentzian shape.
As a result we extract $\alpha \simeq 0.31 \meV/ \mV$, which is in good agreement with the thickness of the barriers and and the spacers. Moreover, using this value of $\alpha$, we check that the LO-phonon replica is at about $36 \pm 5\meV$ after the main resonance peak (see fig.~\ref{fig:iv}b). Thus, $\alpha$ is fairly constant  over the whole bias range.

\begin{figure}
\includegraphics*[width=\linewidth]{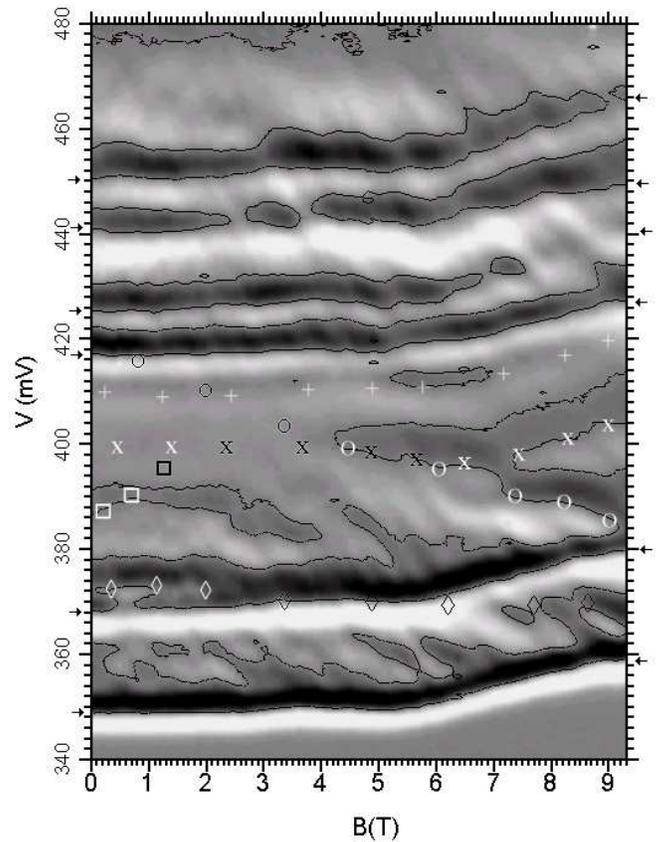}
\caption{Gray-scale plots of the derivative of the differential conductance = $d^{2}I/dV^{2}$, vs bias voltage and magnetic field (steps 0.1mV and 0.02T) numerically obtained from the measured G(V) data. (white, $ dG/dV \geq 5\mu S/\mV$; black, $dG/dV\leq -5 \mu S/\mV$; $dG/dV=0$ corresponds to the contour lines.}
\label{fig:gs}
\end{figure}

We performed  magneto-tunneling measurements  at $T=35mK$ applying  a magnetic field  along the direction parallel to the current flow. Because  the current
 has a sharp increase with the  voltage close to the threshold, we cannot directly plot the conductance data for a quantitative analysis.
In  fig.~\ref{fig:gs}  the derivative of the differential conductance, $d^{2}I/dV^{2}$ is  depicted in a  gray-scale graph. This has no direct physical 
interpretation by itself, but it emphasizes  the peaks of the conductance. 
The  maximum of a conductance peak is represented 
 in this diagram by the boundary line between a clear band and  a dark band, in increasing the  bias. The contour lines represent the $dG/dV=0 \mathrm{ S}.V^{-1}$ values.

This diagram also reveals fluctuations of the Local Density of State (LDoS) of the emitter~\cite{sch97}.   
These fluctuations produce a fan-type pattern  superimposed to the one due to 
the QD states, which becomes  clearly visible at relatively high magnetic field.
 Indeed, for $B \geq 2 T$, the LDoS fluctuations follow the behavior of the Landau levels in the emitter: the slope $\propto | dV |/dB$ of the fluctuations decreases with increasing $B$ and corresponds to Landau levels which are deeply below the emitter Fermi level. We point out that their slope increases whenever a QD ground state is crossed. These slope changes are due to the progressive increase of the electron number in the QD. For $B \leq 2 T$ the Landau levels are destroyed by disorder and the LDoS fluctuations are random. 
Although the LDoS fluctuations are clearly resolved, their current  amplitude is weak compared to the one of the  QD states  and they produce very different features in the gray-scale plot.  Thus, the two phenomena can be distinguished in almost the whole range of $B$ and $V$.  

We focus now on the six clearly resolved boundary lines of the gray-scale, which correspond to the peaks labeled by an arrow in fig~\ref{fig:tdep}. We 
attribute  them  to the N-electron ground states with N=1,2,3,4,5 and 6.
Let us assume a parabolic lateral confining potential $V= \frac{1}{2} m^{*} \omega_{0} r^{2}$ and put 
 the first peak at $ \sim 348 \mV$ in fig.~\ref{fig:gs} in  correspondence 
with  the single particle level  $E_{0,0}$ of the  Darwin-Fock 
spectrum~\cite{foc28} $E_{n,m}= \hbar \omega ( 2 n + |m| + 1 ) + \frac{1}{2} m \hbar \omega_{c}$ (with  $\omega= \sqrt{ \omega_{c}^2 /4 + \omega_{0}^ 2 }$ and $\omega_{c}= e B / m^ {*} $).
 Because  we take  $\alpha = 0.31\meV/\mV$, we obtain $\hbar \omega_{0} = 9 \pm 1 \meV$. This corresponds to a radius 
 $r \simeq \sqrt{2 \hbar/ m^ {*} \omega_{0}} \simeq 15 \pm 1\nm$.
This is fairly larger than the expected radius of a single Si donor in GaAs, which is the most likely expected impurity in the quantum well region. 
On the contrary this value of the radius is in very good agreement with the one extracted from the calculation of the lateral confining potential in the pillar.
 This calculation consists in  solving  the Poisson equation in the Thomas-Fermi approximation for the electronic density in the contacts~\cite{gal92}. The resulting band profile enables us to determine the energy levels of the QD by solving numerically the Sch\"odinger equation. 
The  single particle energy spectrum that is obtained is very well approximated by an harmonic potential with $\hbar \omega_{0} = 7-8 \meV$, and the calculated radius of the ground state is equal to $\simeq 17 \nm$.

Despite the fact that the barriers of our structures have the same thickness, 
we have strong evidence that the second conductance peak ($\sim  368 \mV$ 
in fig.~\ref{fig:gs}) originates from Coulomb blockade. 
The simplest way to include interaction is to assume that the two electron ground state is given by $U_{2}^{0}= 2 E_{0,0} + E_{C}$.
We define  $U_{N}^{n}$ 
as the n-th state of the N-electron spectrum, and $E_{C}$ is the Coulomb interaction:
\begin{equation}
E_{C}= \int
\left| \Psi(\mathbf{r}_{1}, \mathbf{r}_{2}    )   \right| ^ {2}
\frac {e ^ {2} } 
{ 4 \pi \epsilon \left| \mathbf{r}_{2} - \mathbf{r}_{1} \right| }
d\mathbf{r}_{1} d \mathbf{r}_{2}
\label{eq:ec}
\end{equation}
Since we measure the electro-chemical potential of the QD defined by $\mu(N)= U_{N}^{0}-U_{N-1}^{0}$, we interpret the separation between the first and the second peak as $ E_{C}$. 
There is a factor of two discrepancy, however, between 
the experimental value $E_{C}^{exp} = \alpha . 20 \mV \simeq 6-7 \meV$
and the theoretical one estimated perturbatively from 
eq.~\ref{eq:ec} with single particle wavefunctions of the harmonic potential,
 $E_{C}^{th}= e^ {2} \sqrt{ m^{*} \omega_{0} / 32 \pi \hbar \epsilon^{2}} \simeq 14 \meV$. 
 This can be justified by taking into account  the finite width of the QD as well as the screening of  the electrons  in the contacts.
 Following~\cite{bo92}, we estimate the in plane separation of the electrons in the QD as $10 \nm$,
 while the electrons in the contacts are at a distance  
 $\lambda \simeq 13 \nm$ (including the screening length in the contacts $\simeq 8 \nm$).
Therefore,   the Coulomb integral should be cut off 
at separations  $\left| \mathbf{r}_{2} - \mathbf{r}_{1} \right| \geq \lambda$,
what  reduces the bare theoretical Coulomb interaction of  our  device
  to the expected value.

One more argument in favour of our interpretation is the observation that 
  the separation between the two peaks increases slightly with $B$. 
Such an increase has already been observed~\cite{kou97} and is due to the squeezing of the electron orbits which causes an increase of $E_{C}$.

At $B= 0T$, the Darwin-Fock spectrum has degenerate sets of states separated by $\hbar \omega_{0}$, these shells being completely filled for $N=2,6,12 \ldots$. The degeneracy is completely lifted by the electron-electron interaction. 
We attribute the set of four lines in the voltage bias range $420\mV-460\mV$ to the N-electron ground states with N=3,4,5 and 6 which fill progressively the second shell.
 Keeping  $\alpha$ constant in the whole bias range, the experimental separation between the two shells is $\sim 15 \meV$, which is about $ E_C +\hbar \omega_{0}$, as expected in the Constant Interaction (CI) model.
 
\begin{figure}
\includegraphics*[width=\columnwidth]{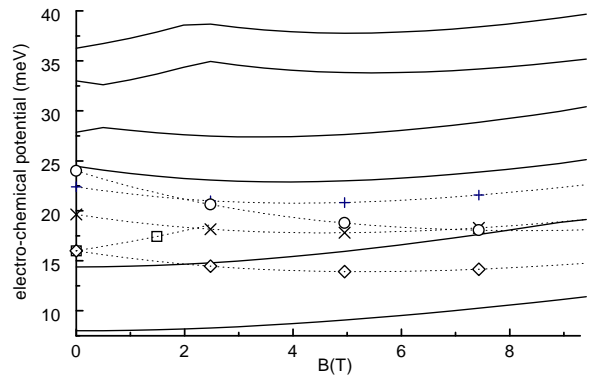}
\caption{ Chemical potential $\mu (N) $ from the calculated energy spectrum. 
The solid lines 
correspond  to  the dot in the ground states  with  $N \leq 6$; the dotted lines correspond to the dot in the lowest  excited states for electron occupancy  
$N=1$  and $2$  (see text for details).}
\label{fig:lanczos}
\end{figure}

Fig.~\ref{fig:lanczos} is  the result of an exact calculation of the energy spectrum of the QD using a Lanczos algorithm, assuming a parabolic potential and a Coulomb interaction ($\hbar \omega_{0}= 8 \meV, E_{C}=6 \meV$).
 The shell structure is clearly visible in  $\mu(N)$, as well as
  the characteristic pairing of levels for particles with opposite spins. 
According to the CI  model, the addition energy  $\mu(N)-\mu(N-1)$ for each 
even electron should always be $E_C $.
  However  $E_{C}^{th}$  is reduced to $5 \meV$  in the second shell, 
because of correlation effects~\cite{ang97} and  we measure an even  smaller 
value $E_{C}^{exp} \simeq 3-4 \meV$ in this shell. 
Again, we attribute the   difference between $E_{C}^{th}$ and $E_{C}^{exp}$ 
mainly to the fact that 
 the assumption of the QD as isolated in the Lanczos calculation is
unphysical. It ignores  the  coupling of the dot with the electrodes.
The occupation rate for the $N \geq 4$ electron ground states appears to 
be noticeably less than that of the first ground states, what implies 
the observed  reduction of the Coulomb blockade.
 Indeed, in the real device, any increase of $V$ lowers the collector barrier height and increases the energy window through which the tunneling occurs, hence enhancing transport  through excited states.

We conclude that there is a close correspondence between the solid lines of 
 fig.~\ref{fig:lanczos} and the six main lines of the experimentally observed 
conductance  diagram and attribute these to tunneling while the dot is in its 
$N-$electron ground state, with increasing $N$.

We claim that the additional temperature dependent features of the fig.~\ref{fig:tdep}, not discussed up to now, are due to the excited states of the N-electron spectra. Such separation of excited and ground states in a non-linear transport experiment implies that the collector barrier is slightly less transparent than the emitter one. This originates probably from
fluctuations of the nominal parameters during the growth.
This hypothesis is corroborated by the data in the negative bias, where the distinction between excited and ground states is not possible.
These peaks appear to have a correspondence to analogous features in the calculated spectrum of 
 fig.~\ref{fig:lanczos}, lying between the first and the second shell.
 The $\lozenge$, $\boxdot$ and $\circ$-dotted lines are respectively the first, second and third one-electron excited states corresponding to the $E_{0,-1}$,$E_{0,1}$ and $E_{0,-2}$ Darwin-Fock states.
 The $\times$ line corresponds to the first excited 2 electrons state $U_{2}^{1}$, with a total spin $S=1$, referred to the single particle ground state energy $U_1^0$ (as all the $U_2^n $ in the following).
 Note that the singlet-triplet transition ($U_{2}^{0}$-$U_{2}^{1}$) occurs for $B \geq 9 T$ and is not observed.
 The $+$ line describes the second 2-electrons excited state $U_{2}^{2}$ with $S=0$. At $B=0T$,  $U_{2}^{2}-U_{2}^{0}= \hbar \omega_{0}$, so that 
 we expect to observe the states in the following order when $V$ increases:
$U_{1}^{0}$,
$U_{2}^{0}$,
$U_{1}^{1}$ (which is 2-fold degenerate),
$U_{2}^{1}$,
$U_{2}^{2}$
and $U_{1}^{2}$.

In fig.~\ref{fig:gs} we have marked the experimental boundary lines that we attribute to these excited states. White symbols refer to lines that are clearly visible in our data. Black symbols indicate the expected position of these lines where the data are confused because of the LDoS fluctuations. Despite these uncertainties, the agreement between experimental and numerical results  is satisfactory. 
We stress that, as in other experiments~\cite{bo92,jou98}, the degeneracy of $E_{0,\pm 1}$ is lifted at $B=0T$. This is a consequence of the lack of inversion symmetry of the Zinc Blende structure. 
Moreover the current contribution of the $E_{0,1}$ washes out quickly with $B$ as expected from the conservation of the angular momentum in 3D-0D tunneling~\cite{jou98,jou98a}.

Let us finally comment on the origin of the main resonance peak observed in the I-V curves (fig.~\ref{fig:iv}b), like in large, non-laterally confined diodes involving tunneling through 2D states. Such a peak appears only if the momentum perpendicular to the interfaces is conserved during the tunneling. This is not the case for transport through a QD where  only the energy is conserved. The only way to explain the shape of the I-V characteristics is to add a continuum of states which  arises from the excited states of the QD.

In a simple non-interacting particles picture there are 12 levels for tunneling (including the 4 ground states of the second shell) when the chemical potential is twice $\hbar \omega_{0}$.
 As the Coulomb interaction suppresses the degeneracy, the mean level spacing is  roughly $\simeq 1 \meV$- which corresponds to our resolution in this experiment. The continuum becomes effective when $\mu \simeq 3 \hbar \omega_{0}$ because of the 64 available states (including the 6 ground states of the third shell); as a consequence there is no way to observe excited levels beyond the second shell of our QD structure.
 
In conclusion, we have reported on  non linear  magneto-tunneling experiment 
conductance  measurements carried out in a single QD. 
Our data are consistent with the spectroscopy of few-electron states, even if LDoS fluctuations decrease the resolution of the experiment. 
We clearly observe a shell structure with a sizeable reduction of the Coulomb 
interaction in the second shell, probably due to screening.
The first excited states for the one- and two- electron states can be identified.
Finally we attribute the main I-V peak to the growing number of excited-states when the voltage bias is increased.
Quantitative discrepancies with exact diagonalization of an isolated QD show that the role of the contacts in these non linear transport experiments cannot 
be ignored.

We would like to thank L.~Couraud for his technical help, M.~Boero, J.P.~Holder and A.K.~Savchenko for fruitful discussions. Work partly supported by 
INFM (Pra97-QTMD) and by EEC with TMR project, contract FMRX-CT98-0180.

\end{document}